\documentclass[twocolumn,showpacs]{revtex4}
\usepackage{tabularx}
\usepackage{dcolumn}
\usepackage{graphics}
\usepackage{bm}
\usepackage[dvips]{graphicx}
\usepackage[dvips]{rotating}
\usepackage[latin1]{inputenc}
\usepackage{dcolumn}
\usepackage{tabularx}
\usepackage{amsmath}
\usepackage{amssymb}
\usepackage{float}
\begin{document}
\draft

\title{Closed-orbit theory for photodetachment in a time-dependent electric field}

\author{B. C. Yang and F. Robicheaux}
\email{robichf@purdue.edu}

\affiliation{Department of Physics and Astronomy, Purdue University,
West Lafayette, Indiana 47907, USA}

\date{\today}

\begin{abstract}
The standard closed-orbit theory is extended for the photodetachment
of negative ions in a time-dependent electric field. The
time-dependent photodetachment rate is specifically studied in the
presence of a single-cycle terahertz pulse, based on exact quantum
simulations and semiclassical analysis. We find that the
photodetachment rate is unaffected by a weak terahertz field, but
oscillates complicatedly when the terahertz pulse gets strong
enough. Three types of closed classical orbits are identified for
the photoelectron motion in a strong single-cycle terahertz pulse,
and their connections with the oscillatory photodetachment rate are
established quantitatively by generalizing the standard closed-orbit
theory to a time-dependent form. By comparing the negative hydrogen
and fluorine ions, both the in-phase and antiphase oscillations can
be observed, depending on a simple geometry of the contributed
closed classical orbits. On account of its generality, the presented
theory provides an intuitive understanding from a time-dependent
viewpoint for the photodetachment dynamics driven by an external
electric field oscillating at low frequency.

\par


\end{abstract}

\maketitle

\section{Introduction}

Quantum effects from closed (or periodic) classical orbits in a
microscopic system have been explored in many different branches of
physics\cite{Gutzwiller}. One of the most typical processes in
atomic and molecular physics is the photoionization of neutral atoms
or the photodetachment of negative ions in external
fields\cite{review}. Its studies often promise an intuitive picture
of the embedded dynamics, which not only reveals an interesting
correspondence between classical and quantum mechanics, but also
allows a better control and manipulation on a microscopic scale. The
general physical picture and formalism are known as closed-orbit
theory\cite{COT01, COT02, COT03} which has been applied or extended
in different situations. However, almost all the systems
investigated before are time independent and therefore energy
conserving\cite{JGao01, JGao02, Peters01, Peters02, Main01, Main02,
Matzkin, Wright, COT_detachment, BCYDD}. Time-dependent systems have
been rarely studied, the one exception being the photoionization of
neutral atoms in a static electric field plus a weak oscillating
field\cite{RFfield, Haggerty}. In this paper, we demonstrate an
application of closed-orbit theory for the photodetachment of
negative ions in a time-dependent electric field.

Many kinds of specific field profiles, like a microwave field or a
low-frequency laser pulse, could be applied to study the
time-dependent effect of an external field on the photodetachment
rate of negative ions. Recently, a strong single-cycle terahertz
(THz) pulse has been available in a table-top experiment. As a
result of its simplicity and other peculiarities, the single-cycle
THz pulse has been applied in exploring the ionization dynamics of
Rydberg atoms\cite{Rydberg01, Rydberg02, Rydberg03}, as well as
controlling the alignment and orientation of polar
molecules\cite{Nelson, Bob}. Inspired in part by these results, we
consider the possibility of using a single-cycle THz field to
manipulate the photodetachment dynamics of negative ions. Temporal
interferences in the time-dependent electron flux (or the
angle-resolved energy spectrum) at large distances were investigated
in a previous paper\cite{BCYFR}, by extending the original idea for
traditional photodetachment microscopy in a static electric
field\cite{Fabrikant, Demkov, Du4983, Blondel}. The classical
trajectory of the photoelectron was tracked from the negative-ion
center to a large distance. We found that some trajectories could
return to the source region when the single-cycle pulse is strong
enough. This observation of closed classical orbits constitutes the
main motivation of this work.

Following the general picture depicted by closed-orbit
theory\cite{COT01, COT02, COT03}, an external field can modulate the
photon absorption rate in the photoionization and photodetachment
processes by driving back an outgoing electron wave to the source
region where the initial bound state is localized. The returning
electron wave interferes with the outgoing wave near the source
center. Each closed classical orbit corresponds to one sinusoidal
term in the total modulation function. Therefore, an oscillatory
photodetachment rate should be expected if the applied single-cycle
THz pulse is strong enough that the electron can be driven back to
the source region. This is indeed observed in our quantum
simulations for a strong THz pulse. In Fig. 1, a representative case
is shown for the negative hydrogen ion (H$^-$). It can be observed
that the photodetachment rate is quite stable in a weak THz pulse
with the maximum field strength $F_m=10kV/cm$, but oscillates in a
complex way when a stronger THz pulse is used such as $F_m=40kV/cm$.
By examining the classical trajectories, no closed orbit is found
for $F_m=10kV/cm$ while, three different types of closed orbits are
found when $F_m=40kV/cm$. These observations are qualitatively
consistent with the general predictions of closed-orbit theory. To
quantitatively understand the oscillatory behavior as in Fig. 1(b),
we have to generalize the existing formulas to include
time-dependent field effects. It turns out that the generalized
formulas agree very well with exact quantum simulations.

\begin{figure}[!t]
\centering
  \includegraphics[width=250pt]{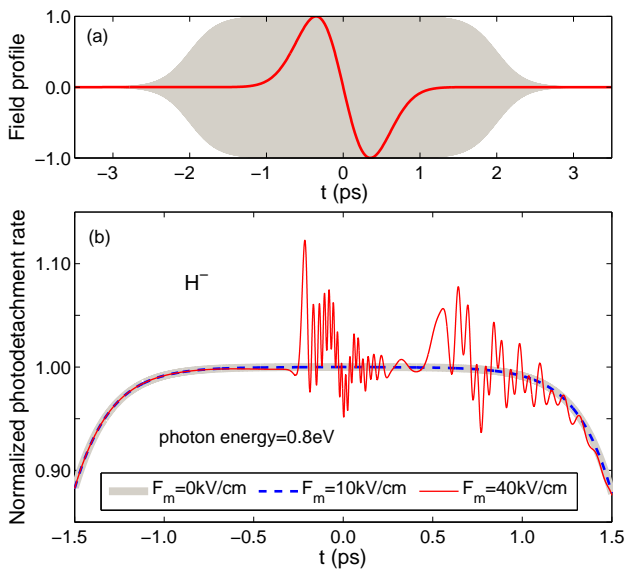}
  \caption{(Color online) (a) Field configurations reproduced from Ref. \cite{BCYFR} with a slightly modification.
  The gray curve and the solid red line represent the weak laser field and the single-cycle THz pulse,
  respectively, divided by their corresponding field amplitudes.
  The laser-field oscillation cannot be resolved due to its high frequency.
  (b) Time-dependent photodetachment rate obtained from exact quantum simulations for H$^-$.
  The THz pulse strengths $F_m$ are given in the legend for different lines, respectively.
  All the curves are normalized using the value of the photodetachment rate
  at $t=0ps$ without the single-cycle pulse applied ($F_m=0kV/cm$).}
\end{figure}

Comparing with the well-established closed-orbit theory for the
photodetachment dynamics in a static electric
field\cite{COT_detachment, BCYDD}, the generalized formulas mainly
have two differences as follows. (a) A static electric field can
always guarantee one and only one closed orbit, and the oscillation
phases are opposite between the photodetachment rates from an
$s$-wave source like negative fluorine ion (F$^-$) and a $p$-wave
source like H$^-$. While, in a time-dependent electric field, no
closed orbit exists if the maximum momentum transfer is not large
enough, but more than one closed orbit may be found if the field is
strong. Determined by a simple property of each closed orbit, both
the in-phase and antiphase oscillations can be observed by comparing
the time-dependent photodetachment rates of H$^-$ and F$^-$. (b) An
electron's kinetic energy is conserved if it was driven back to the
ion by a static field. In contrast, when the electron is driven back
by a time-dependent electric field, its kinetic energy is usually
different from its initial value. Consequently, each sinusoidal term
in the total modulation function is multiplied by an additional
coefficient related to both the electron's outgoing and returning
momenta along the corresponding closed classical orbit.

Although the single-cycle THz pulses are specifically studied, the
presented theory is quite general. The related formulas can be used
directly for the photodetachment of negative ions in any other forms
of the time-dependent electric field, as long as the whole
photodetachment process can be divided approximately into two steps:
one-photon absorption from a weak laser field followed by the
photoelectron motion in the applied external field. The only
additional work needed for a specific system is to identify all the
possible closed classical orbits. We note that an experiment has
been done for the photodetachment of negative chlorine ions in a
microwave field\cite{microwave01, microwave02}, and several
theoretical studies have also been reported\cite{microwave02,
microwave03, microwave04}. However, all of the previous
time-dependent treatments assumed the applied field varied slowly
enough that the electron was driven by a constant electric field
during each detachment event, which approximately corresponds to the
situation in the reported experiment\cite{microwave01, microwave02}.
In this sense, our current work provides further insight into the
general cases of a time-dependent electric field.

In the following section, the theoretical model with a specific
single-cycle THz pulse is briefly summarized, as well as the
numerical method we used for quantum simulations. The three types of
closed classical orbits are identified in Sec. \textrm{III}, and
their corresponding returning waves are specified in Sec.
\textrm{IV}. The general formulas for the time-dependent
photodetachment rate are presented in Sec. \textrm{V} by extending
the standard closed-orbit theory for the photodetachment of negative
ions in a static electric field. Some calculations and discussions
are presented in Sec. \textrm{VI}, followed by a brief conclusion in
Sec. \textrm{VII}. Atomic units are used throughout this work unless
specified otherwise.

\section{Theoretical model and numerical approach}

We choose to study the photodetachment of negative ions (H$^-$ and
F$^-$) in a single-cycle THz pulse as a specific system, based on
several simple reasons as introduced above. Most importantly, this
system has almost all the essential elements expected for the other
general cases, which can be seen clearly in the following sections.
In addition, a numerical solution of Schr\"{o}dinger's equation is
also possible as long as the single-cycle pulse strength is not
extremely large as in Fig. 1(b), which allows us to examine the
accuracy of closed-orbit theory.

The details of the theoretical model has been described in Ref.
\cite{BCYFR}. Here, we give a brief summary and present the
necessary equations related to our present work. We assume the weak
laser field and the applied single-cycle THz pulse are both linearly
polarized along the $z$-axis. The influence of the THz pulse is
negligible on the initial ground state of the negative ion. The much
higher frequency of the laser field relative to the THz pulse allows
the whole photodetachment process to be approximated as two
successive steps: the weakly-bound electron in a short-range
potential well is first released by absorbing one photon from the
weak laser field, and then the photoelectron motion after escaping
from the atom center is mainly guided by the single-cycle THz pulse.

As in our previous paper\cite{BCYFR}, we restrict the weak laser
field within a finite width as in Fig. 1(a). The specific envelope
function has the following form,
\begin{equation}\label{envelope}
f_L(t)=\frac{1}{2}\bigg[\tanh\bigg(\frac{t-t_u}{t_L}\bigg)-\tanh\bigg(\frac{t-t_d}{t_L}\bigg)\bigg]~,
\end{equation}
where $t_d=-t_u=4t_w$ with $t_w$ denoting the single-cycle pulse
duration in the following Eq. (\ref{vector_potential}). The
parameter $t_L$ is selected to be large enough so that the possible
acceleration and deceleration effects are negligible in the outgoing
electron wave when the field envelope is ramping on and off. For
example, the photon energy $\hbar\omega_L$ used for H$^-$ is $0.8eV$
in Fig. 1(b), and $t_L=80T_0$ with $T_0$ approximately $4.6fs$ after
the convention in Ref. \cite{BCYFR}. For F$^-$, the weak-laser field
frequency is chosen to give the same electron kinetic energy $E_0$
as for H$^-$, allowing us to examine effects caused by the different
angular distributions of the initially-outgoing electron waves. For
those laser parameters listed above, the generated outgoing wave at
each initial time $t_i$ can be written as
\begin{equation}\label{initial_wave}
    \psi_0(r,\theta_i,\phi_i, t_i)=f_L(t_i)\psi_{out}(r,\theta_i,\phi_i)e^{-iE_0t_i}~,
\end{equation}
with its amplitude following the laser-field envelope approximately,
where ($r$, $\theta_i$, $\phi_i$) denote spherical coordinates of
the electron relative to the rest atom. The spatial function
$\psi_{out}(r,\theta_i,\phi_i)$ corresponds to the time-independent
outgoing wave generated by a CW laser.

The applied single-cycle THz pulse is assumed to have a
Gaussian-shape vector potential,
\begin{equation}\label{vector_potential}
    A(t)=
    -\frac{F_mt_w}{\sqrt{2}}e^{-\frac{t^2}{t_w^2}+\frac{1}{2}}~,
\end{equation}
which gives a time-dependent single-cycle electric field as in Fig.
1(a) with $F(t)=-dA(t)/dt$. $t_w=0.5~ps$ in Fig. 1, and its value
may be changed for the other calculations. Both the quantum
propagation approach and the semiclassical propagation scheme have
been described in Ref. \cite{BCYFR} for the evolution of the
generated electron wave driven by a single-cycle THz pulse. For a
sufficiently strong THz pulse as in Fig. 1(b), an exact quantum
simulation is possible. The details can be found in Ref.
\cite{BCYFR}, and the basic idea is to solve the following
inhomogeneous Schr\"{o}dinger equation
\begin{equation}\label{inhomogeneous}
    \bigg[i\frac{\partial}{\partial t}-\Big(H_a+H_F(t)-E_0\Big)\bigg]\widetilde{\Psi}(\mathbf{r},t)=f_L(t)D\varphi_i
\end{equation}
on a two-dimensional space spanned by the discretized radial points
and angular momentum basis with different $l$ values. The source
term on the right-hand side of Eq. (\ref{inhomogeneous}) comes from
the interaction of negative ions with a weak laser field, where $D$
and $\varphi_i$ represent, respectively, the dipole operator and the
initial bound state of negative ions. The atomic Hamiltonian
[$\mathbf{p}^2/2+V(r)$] and the interaction term [$F(t)z$] with a
single-cycle THz pulse are denoted, respectively, by $H_a$ and
$H_F(t)$ on the left-hand side of Eq. (\ref{inhomogeneous}). The
specific forms of the binding potential $V(r)$ for H$^-$ and F$^-$
are taken from Ref. \cite{Chu} and Ref. \cite{Lin}, respectively.
The corresponding binding energies $E_b$ are $0.02773$ a.u. and
$0.125116$ a.u. for H$^-$ and F$^-$, respectively, by diagonalizing
the atomic Hamiltonian matrix in a large radial box.

The wave function $\widetilde{\Psi}(\mathbf{r},t)$ in Eq.
(\ref{inhomogeneous}) multiplied by a phase term $\exp(-iE_0t)$ is
the detached-electron wave function at each time instant with a
single-cycle THz pulse applied. Therefore, the time-dependent
photodetachment rate $\Upsilon(t)$ can be calculated
as\cite{Haggerty}
\begin{equation}\label{rate}
    \Upsilon(t)=\frac{d}{dt}\int\widetilde{\Psi}^*(\mathbf{r},t)\widetilde{\Psi}(\mathbf{r},t)d^3\mathbf{r}~.
\end{equation}
In practice, we found that quantum simulations for $\Upsilon(t)$ can
be done efficiently in a smaller radial box than that in Ref.
\cite{BCYFR}, by using a mask function $M(r>r_c)=1-\alpha
[(r-r_c)/(r_m-r_c)]^2\delta t$ to absorb the wave-function part
approaching a large distance after each time step $\delta t$. The
calculation of $\Upsilon(t)$ is done in each time step before the
wave function $\widetilde{\Psi}(\mathbf{r},t)$ multiplied by the
mask function. The absorbing strength $\alpha$, the beginning point
$r_c$ of the mask function and the radial box boundary $r_m$ should
be adjusted carefully to make the numerical results convergent. For
our calculations in this work, we consistently use $\alpha=0.005$.
The appropriate values of $r_c$ and $r_m$ can be chosen by referring
to the classical turning points of the possible closed orbits
discussed in the following section. For instance, $r_c=6500$ a.u.
and $r_m=8000$ a.u. for Fig. 1(b) with $F_m=40kV/cm$. For
$F_m=0kV/cm$ and $F_m=10kV/cm$ in Fig. 1(b), we did not use the mask
function.

\begin{figure}[!t]
\centering
  \includegraphics[width=250pt]{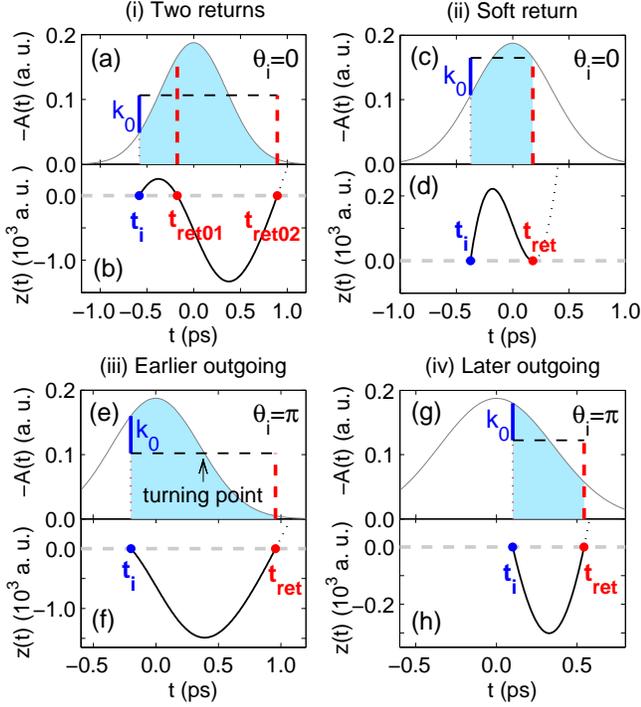}
  \caption{(Color online) Graphic demonstration of the possible closed classical orbits in a strong single-cycle THz pulse.
  The specific case in Fig. 1(b) for $F_m=40kV/cm$ is used as an example.
  The panels titled by (i)-(ii) are two possible cases for $\theta_i=0$, and (iii)-(iv) for $\theta_i=\pi$.
  In each panel, the top subplot is a geometric expression of Eq.
  (\ref{criteria}), and the bottom one shows the corresponding trajectory.
  For each trajectory, the position of the vertical bold solid line (blue online) indicates the starting time while,
  the vertical bold dashed lines (red online) locates its returning time.
  Both the starting and returning time instants are marked accordingly along each closed orbit (solid curve) in (b), (d), (f) and (h),
  where the horizontal bold dashed line shows the ion-center location, with the dotted line representing the continuation of each trajectory.}
\end{figure}

\section{Closed classical orbit}

For our purpose here, we need to find all the possible closed
classical orbits returning back to the atom center. The electron
orbit equation has been obtained as\cite{BCYFR}
\begin{eqnarray}
  \label{rho}
  \rho(t) &=& k_0(t-t_i)\sin(\theta_i)~, \\
  \label{z}
  z(t) &=& [k_0\cos(\theta_i)-A(t_i)](t-t_i)+\int_{t_i}^{t}A(t')dt'~
\end{eqnarray}
in the cylindrical coordinates ($\rho$, $z$) with the atom center at
the origin, where $k_0=\sqrt{2E_0}$. By setting $\rho(t)=0$ and
$z(t)=0$, we get the following condition for the possible closed
orbits:
\begin{equation}\label{criteria}
    [k_0\cos(\theta_i)-A(t_i)](t-t_i)=-\int_{t_i}^{t}A(t')dt'
\end{equation}
with $\theta_i=0$ or $\pi$. This criteria can be expressed
geometrically as in Fig. 2(a), where the rectangular area below the
horizontal dashed line and the shaded area below the reversed
vector-potential curve ($-A(t_i)$) are, respectively, the left- and
right-hand sides of Eq. (\ref{criteria}). The solution to Eq.
(\ref{criteria}) requires an equivalent of these two areas. Note
that the crossings between the horizontal dashed line and the
reversed vector-potential curve correspond to the spatial turning
points ($p_z(t)=0$) of each trajectory, and their corresponding time
is
\begin{equation}\label{turning_0}
    t_<^0=-t_>^0=-t_w\sqrt{\ln\bigg[\frac{A_m}{A(t_i)-k_0}\bigg]}
\end{equation}
for $\theta_i=0$, and
\begin{equation}\label{turning_pi}
    t_>^{\pi}=t_w\sqrt{\ln\bigg[\frac{A_m}{A(t_i)+k_0}\bigg]}
\end{equation}
for $\theta_i=\pi$ with $A_m=-F_mt_w\exp(1/2)/\sqrt{2}$ denoting the
amplitude of the vector potential in Eq. (\ref{vector_potential}).

After a simple geometric analysis in Fig. 2, we can easily determine
the possible range of the starting and returning time for each
closed orbit. In Fig. 2(a) and 2(c) for $\theta_i=0$, the horizontal
dashed line must cross the reversed vector-potential curve, which
requires
\begin{equation}\label{rang_COT01}
\text{$|A(t_i)-A_m|>k_0$ and $t_i<0$, if $\theta_i=0$.}
\end{equation}
Therefore, the starting time $t_i$ of the possible closed orbit must
be negative and less than $-t_w\sqrt{\ln[A_m/(A_m+k_0)]}$. The
corresponding two returning time instants $t_{\mathrm{ret}01}$ and
$t_{\mathrm{ret}02}$ satisfy $t_<^0<t_{\mathrm{ret}01}\leq t_>^0$
and $t_{\mathrm{ret}02}\geq t_>^0$, respectively. In addition, the
position of the second turning point $z(t_>^0)$ cannot be positive
for the trajectory returning back. In Fig. 2(e) and (g) for
$\theta_i=\pi$, the same argument related to the crossing as for
$\theta_i=0$ leads to
\begin{equation}\label{range_COT02}
\text{$-A(t_i)-k_0>0$ and $|A(t)-A_m|>k_0$, if $\theta_i=\pi$.}
\end{equation}
Accordingly, only the classical trajectories starting between
$-t_w\sqrt{\ln(-A_m/k_0)}$ and $t_w\sqrt{\ln(-A_m/k_0)}$ can be
driven back to the source region. The returning time must be
positive and larger than $t_w\sqrt{\ln[A_m/(A_m+k_0)]}$.
Furthermore, the returning time should also be later than the time
of the turning point.

For each closed orbit, the exact starting and returning time can be
found numerically according to Eq. (\ref{criteria}). Figure 3(a)
shows the starting time of each closed orbit as a function of the
corresponding returning instant with $F_m=40kV/cm$. To be clear, we
have categorized all the possible closed orbits into three types
according to their outgoing angle and returning direction. The
first-type closed orbit starts with $\theta_i=0$ and goes back with
$\theta_{\mathrm{ret}}=\pi$, which includes the first-time returned
trajectory in Fig. 2(a)-(b) and the special case shown in Fig.
2(c)-(d). The second-time returned trajectory as in Fig. 2(a)-(b) is
the second-type closed orbit with
$\theta_i=\theta_{\mathrm{ret}}=0$. The other two cases depicted in
Fig. 2(e)-(h) correspond to the third-type closed orbit with
$\theta_i=\pi$ and $\theta_{\mathrm{ret}}=0$. These three types of
closed orbits are distinguished in Fig. 3(a) by the blue solid
curve, the red dotted and the black dashed lines in order. The joint
point between the blue solid curve and the red dotted line in Fig.
3(a) represents a special situation as demonstrated in Fig.
2(c)-(d), which we call a soft return after Refs. \cite{Rost01,
Rost02}. In this case, the atom-center location is just a turning
point of the electron trajectory, and $p_z=0$ when the electron
returns back to the atom center.

\section{Semiclassical returning wave}

For an electron propagating along each classical trajectory, the
corresponding quantum wave can be constructed approximately in a
semiclassical way. To obtain the semiclassical returning wave, we
first choose an initial spherical surface of radius $R$ centered at
the negative ion. As in the standard procedure\cite{COT03,
COT_detachment}, the small radius $R$ is selected near the atom
center such that the initially-outgoing wave is already asymptotic
but not obviously distorted by external fields. Accordingly, the
time-independent function $\psi_{out}(R,\theta_i,\phi_i)$ in Eq.
(\ref{initial_wave}) has the following spherically-outgoing wave
form\cite{Du4983}
\begin{equation}\label{outgoing}
    \psi_{out}(R,\theta_i,\phi_i)=C(k_0)Y_{lm}(\theta_i,\phi_i)\frac{e^{ik_0R}}{R}~,
\end{equation}
on the initial spherical surface. $C(k_0)$ is a complex
energy-dependent coefficient, and $Y_{lm}(\theta_i,\phi_i)$ is a
spherical harmonic function representing the initial angular
distribution of the generated photoelectron wave. For instance,
H$^-$ and F$^-$ considered in this work represent a $p$-wave source
and an $s$-wave source, respectively. The semiclassical wave
corresponding to each trajectory can be written as\cite{Haggerty,
BCYFR}
\begin{equation}\label{semi_wave}
\psi_\nu(t)=f_L(t_i)\psi_{out}(R,\theta_i,\phi_i)\mathcal{A}_\nu
e^{i(\mathcal{S}_\nu-E_0t_i-\lambda_\nu\frac{\pi}{2})}~
\end{equation}
where the subscript $\nu$ labels the considered trajectory.
$\mathcal{A}$ and $\mathcal{S}$ denote, respectively, the
semiclassical amplitude and classical action accumulated in an
augmented phase space. The Maslov index $\lambda$ is determined by
the number of singularities in $\mathcal{A}$ along the trajectory.

When the electron wave returns back to the source region, the
returning wave behaves like a plane wave
approximately\cite{COT_detachment, BCYDD}, traveling along the
returning direction of the closed classical orbit. Therefore, by
connecting the semiclassical wave in Eq. (\ref{semi_wave}) to a
plane wave form, the electron returning wave along each closed orbit
can be approximated as
\begin{equation}\label{returning_wave}
    \psi^{\nu}_{\mathrm{ret}}(t)=f_L(t_i)e^{-iE_0t}\widetilde{\psi}^{\nu}_{\mathrm{ret}}
\end{equation}
with the reduced function $\widetilde{\psi}^{\nu}_{\mathrm{ret}}$
specifically expressed as
\begin{equation}\label{wave_ret_I}
\widetilde{\psi}^\mathrm{I}_{\mathrm{ret}}=C(k_0)\mathcal{G}_{\mathrm{co}}Y_{lm}(\theta_i=0)e^{-ik_{\mathrm{ret}}z}~,
\end{equation}
\begin{equation}\label{wave_ret_II}
\widetilde{\psi}^\mathrm{II}_{\mathrm{ret}}=C(k_0)\mathcal{G}_{\mathrm{co}}Y_{lm}(\theta_i=0)e^{ik_{\mathrm{ret}}z}~,
\end{equation}
\begin{equation}\label{wave_ret_III}
\widetilde{\psi}^\mathrm{III}_{\mathrm{ret}}=C(k_0)\mathcal{G}_{\mathrm{co}}Y_{lm}(\theta_i=\pi)e^{ik_{\mathrm{ret}}z}~,
\end{equation}
for the wave parts returned along the three types of closed orbits
as in Fig. 3(a), respectively. Note that the returning electron
momentum $k_{\mathrm{ret}}$ in Eq.
(\ref{wave_ret_I})-(\ref{wave_ret_III}) is generally different from
the initial momentum $k_0$ when a time-dependent electric field is
applied, which is different from the case of a static field. In the
above equations,
\begin{equation}\label{Ylm}
    Y_{lm}(\theta_i=\pi)=(-1)^lY_{lm}(\theta_i=0)=(-1)^lN_{l0}\delta_{m0}~,
\end{equation}
with $N_{l0}=\sqrt{(2l+1)/(4\pi)}~$, and the factor
\begin{equation}\label{Gco_reduced}
    \mathcal{G}_{\mathrm{co}}=\frac{\mathcal{A}}{R}e^{i(\widetilde{\mathcal{S}}-\lambda\frac{\pi}{2})}
\end{equation}
representing the wave amplitude and phase accumulated along each
closed orbit. The redefined action function
$\widetilde{\mathcal{S}}$ in Eq. (\ref{Gco_reduced}) has the
following form
\begin{equation}\label{extended_action}
    \widetilde{\mathcal{S}}=\mathcal{S}+E_0(t-t_i)~,
\end{equation}
which is called an ``extended action'' in Ref. \cite{Haggerty}.

\begin{figure}[!t]
\centering
  \includegraphics[width=250pt]{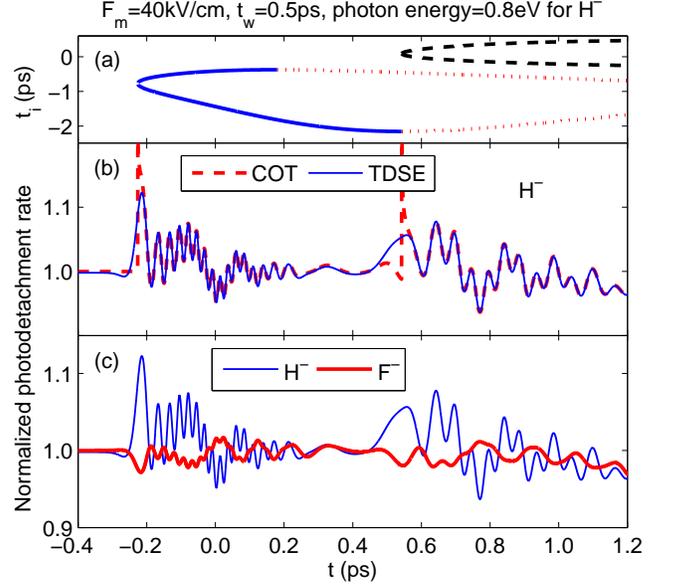}
  \caption{(Color online) (a) Returning time plot for the relevant closed orbits in Fig. 1(b) and
  Fig. 2, giving the initial, outgoing time $t_i$ for each possible
  closed orbit returning back to the atom center at time $t$.
  The three types of closed orbits are identified in order by the solid blue curve, the dotted and the dashed
  lines.
  (b) Time-dependent photodetachment rate for H$^-$. The bold dashed curve is given
  by Eqs. (\ref{modulation_sum}), (\ref{modulation_nu}) and (\ref{Ck}) from closed-orbit theory (COT),
  while, the solid blue curve is from exact quantum simulations
  by directly solving the time-dependent Schr\"{o}dinger equation (TDSE) in Eq. (\ref{inhomogeneous}).
  (c) Comparison between the photodetachment rates for H$^-$ (thin and blue curve) and F$^-$ (bold and red line), obtained from quantum simulations.
  All the quantum results have been normalized
  using the value of the photodetachment rate at $t=0ps$ without any external fields applied.}
\end{figure}

The general expressions for the semiclassical amplitude
$\mathcal{A}$ and classical action $\mathcal{S}$ have been obtained
in Ref. \cite{BCYFR} for the electron wave propagating along any
classical trajectories in a time-dependent external field. For the
closed classical orbits involved here, we have
\begin{equation}\label{amplitude01}
    \frac{\mathcal{A}}{R}=\frac{1}{k_0(t-t_i)}\bigg|\frac{k_0}{k_0-F(t_i)(t-t_i)\cos(\theta_i)}\bigg|^{1/2}
\end{equation}
from Eq. (28) in Ref. \cite{BCYFR} with $\theta_i=0$ or $\pi$, which
can be rewritten as an intuitive form (Appendix A)
\begin{equation}\label{amplitude02}
    \frac{\mathcal{A}}{R}=\frac{1}{k_0(t-t_i)}\bigg|\frac{p_z(t_i)dt_i}{p_z(t)dt}\bigg|^{1/2}
\end{equation}
with $p_z(t_i)$ and $p_z(t)$ denoting, respectively, the outgoing
and returning momenta of the corresponding closed orbit. Note that
the absolute-square-root part on the right-hand side of Eqs.
(\ref{amplitude01}) and (\ref{amplitude02}) becomes a unity when the
energy is conserved, and Eqs. (\ref{amplitude01}) and
(\ref{amplitude02}) reduce to a static-field case as in
\cite{COT_detachment, BCYDD}. Based on Eq. (\ref{amplitude02}), the
Maslov index $\lambda$ can be determined from the returning-time
plot as in Fig. 3(a). For example, if
$\theta_{\mathrm{ret}}=\theta_i=0$ as the second-type closed orbit
in Eq. (\ref{wave_ret_II}), $\lambda=0$ when the slope $dt/dt_i$ in
Fig. 3(a) is positive, otherwise $\lambda=1$. From Eq. (B9) in Ref.
\cite{BCYFR}, the ``extended action'' $\widetilde{\mathcal{S}}$ in
Eq. (\ref{modulation_nu}) can be obtained as (Appendix A)
\begin{equation}\label{extended_action_expression}
    \widetilde{\mathcal{S}}=\Big[k_0^2+\frac{1}{2}A^2(t_i)-A(t_i)k_0\cos(\theta_i)\Big](t-t_i)-\frac{1}{2}\int_{t_i}^tA^2(t')dt'
\end{equation}
by using the condition in Eq. (\ref{criteria}) for a trajectory
returning back to the source region.

\section{Closed-orbit theory}

Following the general picture established in Refs. \cite{COT01,
COT02, COT03, Haggerty}, the total photodetachment rate $\Upsilon
(t)$ in Eq. (\ref{rate}) can be decomposed as
\begin{equation}\label{rate_sum}
    \Upsilon(t)=\Upsilon_0(t)+\sum_\nu\Upsilon_\nu(t)
\end{equation}
where
\begin{equation}\label{rate_00}
    \Upsilon_0(t)=-2~\text{Im}\langle I(t)|\psi_{\mathrm{dir}}(t)\rangle
\end{equation}
is a smooth background representing the photodetachment rate without
any external fields, and
\begin{equation}\label{rate_nu}
    \Upsilon_\nu(t)=-2~\text{Im}\langle I(t)|\psi^{\nu}_{\mathrm{ret}}(t)\rangle
\end{equation}
is contributed by the returning electron wave
$\psi^{\nu}_{\mathrm{ret}}(t)$ associated with the $\nu$-th closed
orbit. The wave-source function $I(t)$ in Eqs. (\ref{rate_00}) and
(\ref{rate_nu}) has the following form
\begin{equation}\label{source_term}
    I(t)=f_L(t)e^{-iE_0t}D\varphi_i
\end{equation}
which is the source term on the right-hand side of Eq.
(\ref{inhomogeneous}) multiplied by a phase term $\exp(-iE_0t)$. The
smooth background in Eq. (\ref{rate_00}) can be worked out as
\begin{equation}\label{background}
    \Upsilon_0(t)=f^2_L(t)k_0|C(k_0)|^2
\end{equation}
according to the existing formulas in Ref. \cite{BCYDD} after in Eq.
(\ref{rate_00}) using the directly-outgoing wave
$\psi_{\mathrm{dir}}(t)$ given by Eq. (\ref{initial_wave}) and the
wave-source function $I(t)$ in Eq. (\ref{source_term}).

To calculate the contributed term from the returning electron wave,
we first rewrite Eq. (\ref{rate_nu}) as
\begin{equation}\label{rate_adapt}
    \Upsilon_\nu(t)=-2f_L(t)f_L(t_i)\text{Im}\langle D\varphi_i|\widetilde{\psi}^{\nu}_{\mathrm{ret}}\rangle
\end{equation}
using the expressions of the returning wave in Eq.
(\ref{returning_wave}) and the wave source function in Eq.
(\ref{source_term}). The overlap integration in Eq.
(\ref{rate_adapt}) between a static wave source $D\varphi_i$ and a
reduced returning wave $\widetilde{\psi}^{\nu}_{\mathrm{ret}}$ is
now in a familiar form usually encountered when a static field is
applied. Accordingly, the same manipulations as in Ref. \cite{BCYDD}
can be followed, and the final expression is (Appendix B)
\begin{equation}\label{overlap_integration}
    \text{Im}\langle
    D\varphi_i|\widetilde{\psi}^{\nu}_{\mathrm{ret}}\rangle=\frac{\textsl{g}^l}{2}(2l+1)\delta_{m0}\text{Im}\big[C^*(k_0)C(k_{\mathrm{ret}})\mathcal{G}^*_{\mathrm{co}}\big]
\end{equation}
where $\textsl{g}=1$ for a closed orbit with the same outgoing and
returning directions as in Eq. (\ref{wave_ret_II}), and
$\textsl{g}=-1$ for a closed orbit with opposite outgoing and
returning directions as in Eqs. (\ref{wave_ret_I}) and
(\ref{wave_ret_III}). Therefore, the sinusoidal term
$\Upsilon_\nu(t)$ in Eq. (\ref{rate_sum}) corresponding to each
closed orbit is obtained as
\begin{eqnarray}
\nonumber
    \Upsilon_\nu(t) &=& \textsl{g}^lf_L(t)f_L(t_i)(2l+1)\delta_{m0}~~\\
    \label{sinusoidal_term}
    && \times C^*(k_0)C(k_{\mathrm{ret}})\frac{\mathcal{A}}{R}\sin\Big(\widetilde{\mathcal{S}}-\lambda\frac{\pi}{2}\Big)
\end{eqnarray}
by substituting Eq. (\ref{overlap_integration}) into Eq.
(\ref{rate_adapt}). To write down Eq. (\ref{sinusoidal_term}), we
have assumed that the energy-dependent coefficient $C(k)$ is either
a real function or a pure imaginary function as discussed in Refs.
\cite{Bracher_exact, Bracher_thesis}.

Finally, the closed-form expression for the photodetachment rate can
be written as a product like
\begin{equation}\label{rate_product}
    \Upsilon(t)=\Upsilon_c\mathcal{H}(t)
\end{equation}
by combining Eqs. (\ref{rate_sum}), (\ref{background}) and
(\ref{sinusoidal_term}) together, where
\begin{equation}\label{rate_CW}
    \Upsilon_c=k_0|C(k_0)|^2
\end{equation}
representing the photodetachment rate in a weak CW laser field with
$f_L(t)=1$, and
\begin{equation}\label{modulation_sum}
    \mathcal{H}(t)=f^2_L(t)+\sum_\nu f_L(t)f_L(t_i^\nu)\mathcal{H}^\nu(t)
\end{equation}
containing both the possible effects induced by an external field
and the slowly-varying envelope $f_L(t)$ assumed for the weak laser
field. In the summation of Eq. (\ref{modulation_sum}), one of the
laser-field envelope function $f_L(t)$ is evaluated at $t_i$, which
was brought in by the initially-outgoing wave in Eq.
(\ref{initial_wave}) through the returning wave in Eq.
(\ref{returning_wave}). The presence of the prefactor
$f_L(t)f_L(t_i)$ in Eq. (\ref{modulation_sum}) requires the
occurrence of the laser excitations at both the instant $t_i$ and
$t$, as well as a quantum coherence between these two excitation
events. $\mathcal{H}(t)$ in Eq. (\ref{modulation_sum}) is usually
called a modulation function of the photodetachment rate, and
$\mathcal{H}^\nu(t)$ corresponds to the contribution from each
closed orbit, which can be explicitly written as
\begin{equation}\label{modulation_nu}
    \mathcal{H}^\nu(t)=\textsl{g}^l(2l+1)\delta_{m0}\frac{C(k_{\mathrm{ret}})}{C(k_0)}\frac{\mathcal{A}}{k_0R}\sin\Big(\widetilde{\mathcal{S}}-\lambda\frac{\pi}{2}\Big)
\end{equation}
from Eqs. (\ref{sinusoidal_term})-(\ref{modulation_sum}). Required
by a Wigner power law such as $\Upsilon_c\propto(k_0)^{2l+1}$ near
the photodetachment threshold\cite{Wigner}, the energy-dependent
coefficient $C(k)$ is proportional to $k^l$ for small $k$. For H$^-$
especially, an analytic form of $C(k)$ needed in Eq.
(\ref{modulation_nu}) has been given by a well-established model in
Refs. \cite{Du4983, Du5609}, and
\begin{equation}\label{Ck}
    \frac{C(k_{\mathrm{ret}})}{C(k_0)}=\frac{k_{\mathrm{ret}}(E_b+E_0)^2}{k_0(E_b+E_{\mathrm{ret}})^2}
\end{equation}
with $E_{\mathrm{ret}}=k_{\mathrm{ret}}^2/2$. For $E_0\ll E_b$ and
$E_{\mathrm{ret}}\ll E_b$, Eq. (\ref{Ck}) reduces to that given by
the Wigner threshold law.

At the end of this section, we would like to point out that no
specific profile has been assumed for the applied electric field in
the above derivation. All the formulas obtained in this section are
generally applicable for the photodetachment in a time-dependent
electric field oscillating at low frequency, which automatically
includes the static-field case as in Refs. \cite{COT_detachment,
BCYDD}. Since the semiclassical propagation scheme and its related
formulas in Ref. \cite{BCYFR} were also presented in a general form
for the electron propagating along any classical trajectories, the
established theory, with Ref. \cite{BCYFR} together, provides an
intuitive understanding for the photodetachment dynamics driven by a
slowly-varying oscillating electric field, which depicts a clear
dynamical picture from a time-dependent viewpoint based on the
classical trajectory propagation.

\section{Calculations and discussions}

In this section, we present some specific calculations using Eqs.
(\ref{modulation_sum}) and (\ref{modulation_nu}) for the
photodetachment driven by a single-cycle THz pulse as illustrated in
Fig. 1(a), as well as the related discussions on the possible
interesting effects expected from the current theory. The exact
quantum simulations are also presented for several cases to examine
the accuracy of the simple formulas in Eqs. (\ref{modulation_sum})
and (\ref{modulation_nu}). For H$^-$ as a $p$-wave source, the
expression in Eq. (\ref{Ck}) is used. For F$^-$ as an $s$-wave
source, we consistently use $C(k_{\mathrm{ret}})/C(k_0)=1$ after the
Wigner power law near the photodetachment threshold.

\begin{figure}[!b]
\centering
  \includegraphics[width=250pt]{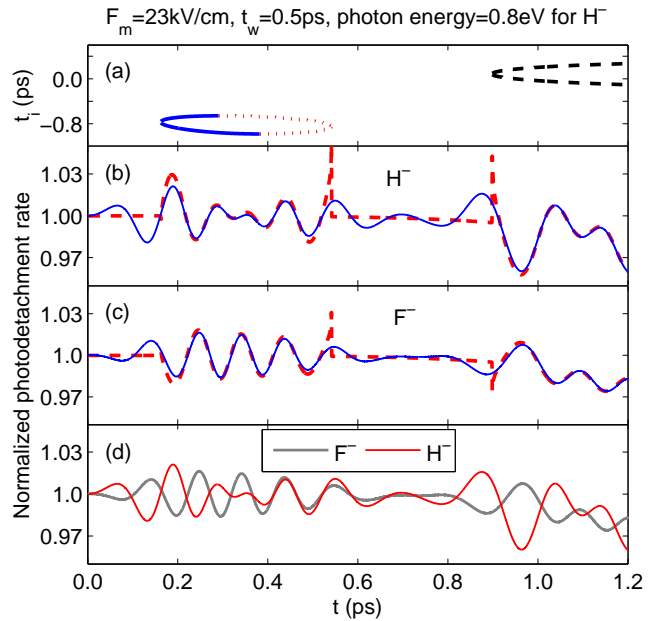}
  \caption{(Color online) Phase dependence on a simple geometry of the closed orbit. (a) Returning time plot for the relevant closed orbits.
  The solid blue curve, the dotted and the dashed lines correspond
 to the closed orbits with ($\theta_i=0$, $\theta_{\mathrm{ret}}=\pi$),
 $\theta_i=\theta_{\mathrm{ret}}=0$ and ($\theta_i=\pi$,
 $\theta_{\mathrm{ret}}=0$), respectively.
  (b)-(c) give the time-dependent photodetachment rate for H$^-$ and F$^-$, respectively.
  The bold dashed and the solid blue curves are given by closed-orbit theory and quantum simulations, respectively.
  (d) Comparison between the photodetachment rates for H$^-$ (thin and red curve) and F$^-$ (bold and gray line), obtained from quantum simulations.
  All the quantum results have been normalized using the value of the photodetachment rate at $t=0ps$ without any
  external fields applied.}
\end{figure}

\begin{figure}[!t]
\centering
  \includegraphics[width=250pt]{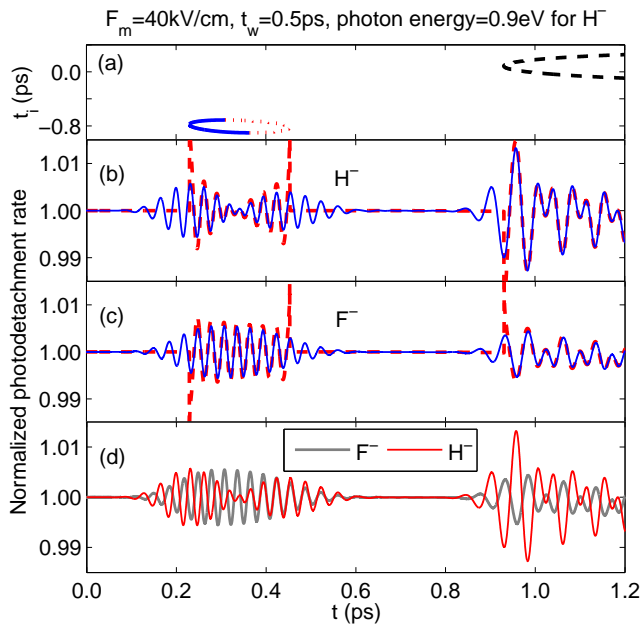}
  \caption{(Color online) The same as Fig. 4 but for higher photon energy and larger field strength as listed on the top.
  Here, $t_L$ in Eq. (\ref{envelope}) was set to be $50T_0$ as in Ref. \cite{BCYFR}.}
\end{figure}

\begin{figure}[!t]
\centering
  \includegraphics[width=250pt]{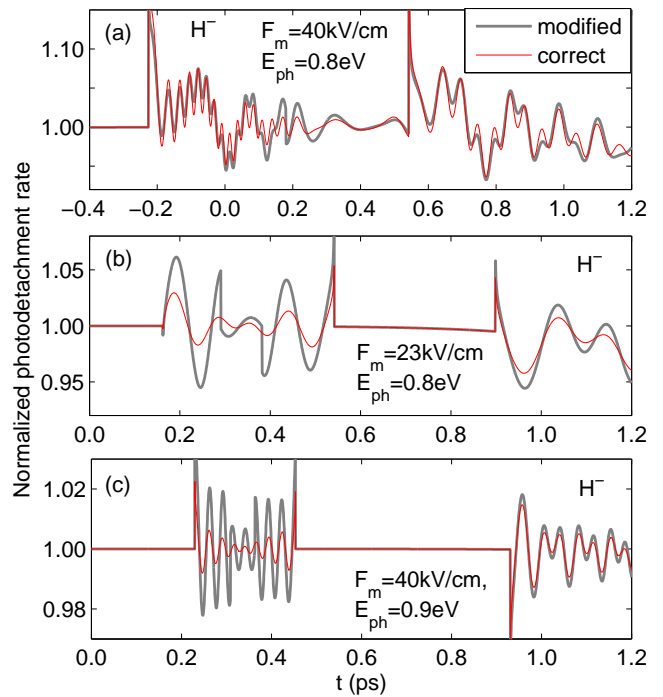}
  \caption{(Color online) Amplitude dependence on the returning electron momentum for H$^-$.
  The thin solid and red curves in (a), (b) and (c) are reproduced from the bold dashed lines in Figs. 3(b), 4(b) and 5(b), respectively, for comparison.
  The bold gray lines are calculated from Eq. (\ref{modulation_nu}) without the term $C(k_{\mathrm{ret}})/C(k_0)$ included.}
\end{figure}

\begin{figure}[!t]
\centering
  \includegraphics[width=250pt]{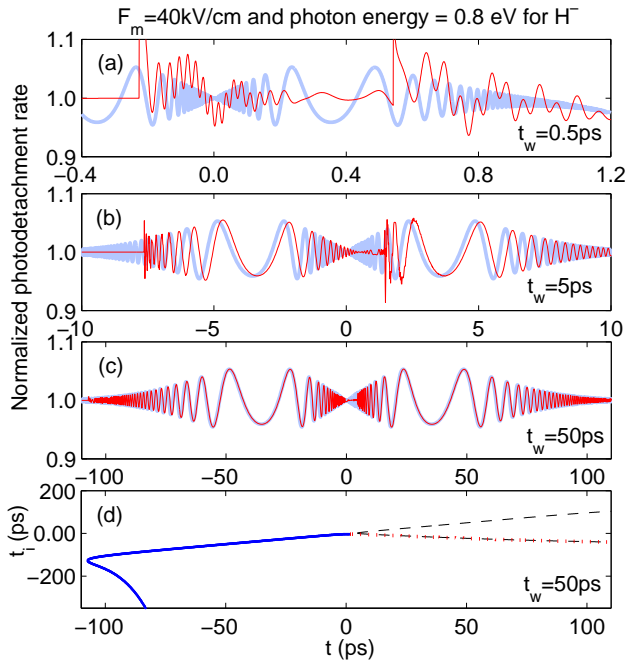}
  \caption{(Color online) (a)-(c) Comparison between the time-dependent
  calculations from Eq. (\ref{modulation_sum}) (heavy thin curves)
  and a static-field approximation after Eq. (\ref{static_rate}) (light bold lines) by increasing the single-cycle pulse duration.
  (d) The returning time plot for the relevant closed orbits in (c).
  The solid blue curve, the dotted and the dashed lines correspond
 to the closed orbits with ($\theta_i=0$, $\theta_{\mathrm{ret}}=\pi$),
 $\theta_i=\theta_{\mathrm{ret}}=0$ and ($\theta_i=\pi$,
 $\theta_{\mathrm{ret}}=0$), respectively.}
\end{figure}

\begin{figure}[!t]
\centering
  \includegraphics[width=250pt]{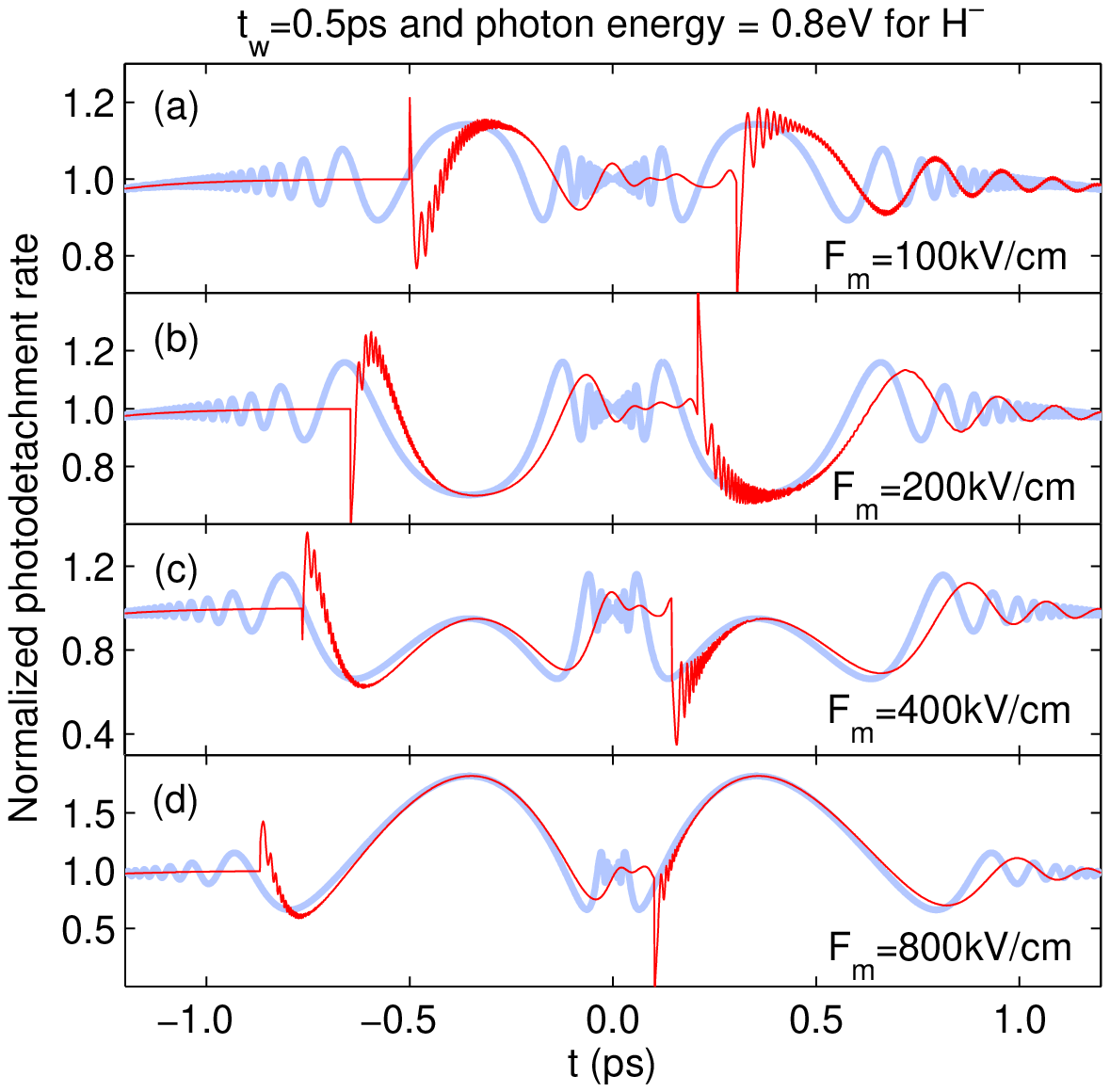}
  \caption{(Color online) Comparison between the time-dependent
  calculations from Eq. (\ref{modulation_sum}) (heavy thin curves)
  and a static-field approximation after Eq. (\ref{static_rate}) (light bold lines) by increasing the single-cycle pulse strength.}
\end{figure}

The photodetachment rate given by the generalized closed-orbit
theory is compared with that from exact quantum simulations in Fig.
3(b) by taking the demonstrated case in Fig. 1 (b) for instance.
Excellent agreement can be found in the time range where the
electron can be classically driven back to the source region. Near
the classical boundary corresponding to the left-most points of the
solid blue line and the dashed black curve in Fig. 3(a), the results
from closed orbit theory diverge as usual because $dt/dt_i=0$ in the
semiclassical amplitude as in Eq. (\ref{amplitude02}). Besides, as
shown by the quantum simulations in Figs. 3-5, the oscillatory
behavior of the photodetachment rate can also be observed in the
classically forbidden range where no closed classical orbit was
found, and the oscillation amplitude goes to zero gradually with
time $t$ away from the classical boundary.

The above mentioned oscillation behavior in the classically
forbidden region looks similar to that observed in Ref.
\cite{BCYbarrier} for a static barrier, where the continuation of
the cross-section oscillations beyond the standard closed-orbit
theory was explained as an effect of quantum over-barrier
reflection. However, our current time-dependent system is more
complicated than the static case in Ref. \cite{BCYbarrier}. To
quantitatively describe those extended oscillations in the
classically forbidden region, as well as to repair the divergence
near the classical boundary as in Fig. 3(b), some sophisticated
manipulations are needed beyond our current treatment based on real
classical trajectories\cite{Main02, BCYDD, BCYbarrier}.
Nevertheless, the semiclassical formulas obtained in Sec. \textrm{V}
are already insightful enough to understand the physics behind the
complicated oscillations of the photodetachment rate as in Fig.1
(b). Therefore, in the following discussion, we will mainly focus on
the physics revealed by Eqs. (\ref{modulation_sum}) and
(\ref{modulation_nu}), instead of pursuing an appropriate way to
save the semiclassical description near the classical boundary and
in the classically forbidden region. In addition, it can be seen
from Figs. 3-5 that quantum effects in the classically forbidden
region become negligible when the THz pulse gets stronger relative
to the initial electron kinetic energy.

Beyond the specific case of H$^-$, the established theory in Sec.
\textrm{V} promises an interesting discrepancy between the
oscillation behaviors of the photodetachment rates for different
negative ions. To give a general impression, the photodetachment
rate for F$^-$ is also displayed in Fig. 3(c) from quantum
calculations, where both the oscillation amplitude and phase are
very different from those observed for H$^-$. Since the oscillations
in Fig. 3 are too complicated to give any clear information, we
choose to first examine the two simpler cases in Figs. 4 and 5,
where the time ranges corresponding to the three types of closed
orbits are almost separated between each other, supported
qualitatively by the graphic analysis as in Fig. 2 with Eqs.
(\ref{turning_0}) and (\ref{turning_pi}) together. Comparing with
those parameters in Fig. 3, the field strength is smaller in Fig. 4
while the photon energy is higher in Fig. 5. For both the two cases
in Figs. 4 and 5, a good agreement between closed-orbit theory and
quantum calculations can also be found in the classically allowed
range for closed classical orbits. By comparing H$^-$ and F$^-$ in
each case as in Figs. 4(d) and 5(d), different phase relations can
be observed in different time ranges for the oscillations of the
photodetachment rates.

According to Eq. (\ref{modulation_nu}) for each specific closed
orbit, the modulation phase for different negative ions is only
determined by a $\textsf{g}$-coefficient associated with a simple
geometry of the closed orbit. If the contributed closed orbit has
the opposite outgoing and returning directions as those identified
by the solid and the dashed curves in Figs. 3(a), 4(a) and 5(a), the
pre-factor $\textsf{g}^l$ can be $1$ or $-1$, depending on whether
the quantum number $l$ is even or odd. Therefore, the
photodetachment-rate for H$^-$ ($p$-wave source) oscillates out of
phase with that for F$^-$ ($s$-wave source) in the corresponding
time ranges. In contrast, if the outgoing and returning directions
of the closed orbit are the same like the second-type closed orbit
indicated by the dotted lines in Figs. 3(a), 4(a) and 5(a), then the
pre-factor $\textsf{g}^l=1$ for all the $l$ values. Accordingly, the
photodetachment rates for H$^-$ and F$^-$ oscillate in phase in the
time range of the second-type closed orbit. These arguments based on
the $\textsf{g}$-coefficient in Eq. (\ref{modulation_nu})
successfully explain the different phase relations between the
oscillatory curves for H$^-$ and F$^-$ in different time ranges of
Figs. 4(d) and 5(d).

More interestingly, near $t\sim0.3ps$ in both Figs. 4 and 5 where
time ranges for the first- and second-type closed orbits overlap,
the oscillation amplitude of the photodetachment rate for H$^-$
becomes smaller but the amplitude for F$^-$ gets larger. This is
also caused by the $\textsf{g}$-coefficient. For H$^-$,
$\textsf{g}^l=-1$ and $1$ for the first- and second-type closed
orbits, respectively. Therefore, the total contributions from these
two closed orbits at each time becomes smaller because of a
large-part cancelation between each other in Eq.
(\ref{modulation_sum}). For F$^-$, both types of closed orbits have
$\textsf{g}^l=1$, and accordingly their incoherent summation in Eq.
(\ref{modulation_sum}) makes the oscillation amplitude larger.
Nevertheless, the same phenomenon does not appear in Fig. 3(c),
which cannot be simply explained by the $\textsf{g}$-coefficient.
Note that the closed orbits starting near $t_i=-2ps$ need a much
longer time to be driven back to the source region than those
starting near $t_i=0ps$ do. Consequently, their contributions in Eq.
(\ref{modulation_sum}) are negligible compared with those
contributed by the closed orbits starting much later, as a result of
their associated much weaker amplitudes $\mathcal{A}$ according to
Eq. (\ref{amplitude01}). This can be seen clearly in the overlap
range near $t=0.4ps$ in Fig. 3(c), where the oscillation behavior is
dominated by the second-type closed orbit, and the photodetachment
rates for H$^-$ and F$^-$ oscillate in phase.

Following Eq. (\ref{modulation_nu}), the modulation amplitude is
generally affected by both the energy-dependent coefficient $C(k)$
and the angular distribution of an initially outgoing wave.
Specifically, for an $s$-wave source like F$^-$, the effect of
$C(k)$ is negligible according to the Wigner power law near the
photodetachment threshold. However, for a $p$-wave source like
H$^-$, the effect of $C(k)$ is not negligible in principle according
to Eqs. (\ref{modulation_nu}) and (\ref{Ck}), and the change of the
electron returning momentum $k_{\mathrm{ret}}$ relative to the
initially-outgoing momentum $k_0$ might modify the oscillation
amplitude of the photodetachment rate dramatically. As a
demonstration, the correct semiclassical results in Figs. 3(b), 4(b)
and 5(b) are compared in Figs. 6(a)-6(c), respectively, with their
modified calculations according to Eq. (\ref{modulation_nu}) without
the term $C(k_{\mathrm{ret}})/C(k_0)$ included. Note that the
modified results give an oscillation amplitude only determined by
the factor $(2l+1)$ in Eqs. (\ref{modulation_sum}) and
(\ref{modulation_nu}) for the same types of closed orbits. By
comparing Figs. 6(a)-6(c) and Figs. 3(c), 4(d) and 5(d)
correspondingly, we can conclude that the oscillation-amplitude
discrepancy between the photodetachment rates for H$^-$ and F$^-$ in
Fig. 3(c) is mainly caused by the different $l$ values associated
with the wave source property, while, the almost equal oscillation
amplitudes observed in both Figs. 4(d) and 5(d) are induced by the
change of the returning momentum related to the external
time-dependent field.

A dynamic picture of the electron-momentum variation and also the
difference between $k_{\mathrm{ret}}$ and $k_0$ can be obtained
visually from a graphic demonstration of Eq. (\ref{criteria}) as in
Fig. 2. The vertical vector from the reverted vector-potential curve
to the horizontal dashed line as in Fig. 2(a) is just the electron
momentum vector at each time instant. The specific case illustrated
in Fig. 2 corresponds to Fig. 3 and Fig. 6(a). There is no big
difference observed between $k_{\mathrm{ret}}$ and $k_0$ in Fig. 2
for most cases, especially for the first-time returning orbit in
Fig. 2(a) and those closed orbits as in Fig. 2(g) which contribute
dominantly to the final oscillation amplitude because of their short
durations. This explains the small discrepancy between the modified
results and the correct calculations in Fig. 6(a). Besides the
overall agreement, an obvious discrepancy can be found near
$t=0.2ps$ in Fig. 6(a). This is because the electron experiences a
soft return as in Fig. 2(c) where the returning momentum is zero.
Accordingly, as shown in Fig. 6(a), the oscillation amplitude from
correct calculations appears smaller than that given by the modified
results without the term $C(k_{\mathrm{ret}})/C(k_0)$ included in
Eq. (\ref{modulation_nu}).

There are mainly three differences between the cases in Figs. 6(b)
and 6(c) and that in Fig. 6(a). First, all the closed orbits almost
contribute equally to the oscillation amplitudes in Figs. 6(b) and
6(c) as a result of their similar durations. Second, compared with
those parameters in Fig. 6(a), the field strength is weaker in Fig.
6(b) and the photon energy is higher in Fig. 6(c). If we make a
corresponding change in Fig. 2 after Fig. 4(a) or Fig. 5(a), the
obtained returning momenta are almost always smaller than the
initial values. This is the main reason why the modified results
give a larger amplitude than the correct calculations in Figs. 6(b)
and 6(c). The third important difference is that the time range for
the first- and second-type closed orbits in Figs. 4(a) and 5(a) is
much more localized near the time instant of a soft return than that
in Fig. 3(a). This is another origin for the large discrepancy
observed near $t\sim 0.4ps$ in Figs. 6(b) and 6(c). An additional
interesting effect related to the soft return is the discontinuity
of the modified results as in Fig. 6, which is caused by the sign
change of $\textsf{g}^l$ on the two sides of the softly-returning
time.

Another interesting aspect related to our present work is about the
static-field approximation in a long wave-length limit of the
applied oscillating field. As introduced in Sec. \textrm{I}, it has
been found that the static-field approximation for each
photodetachment event works very well for an experiment in a strong
microwave field\cite{microwave01, microwave02}. Our current theory
has already allowed us to examine this simple quasi-static picture
from a time-dependent viewpoint. For this purpose, we take H$^-$ for
instance, and compare the time-dependent photodetachment rate given
by Eq. (\ref{modulation_sum}) with the quasi-static result obtained
from\cite{COT_detachment, BCYDD}
\begin{equation}\label{static_rate}
    \mathcal{H}^F(t)=f^2_L(t)\bigg[1+\frac{1}{\mathcal{S}_F}\cos(\mathcal{S}_F)\delta_{m0}\bigg]~,
\end{equation}
where $\mathcal{S}_F=4\sqrt{2}E_0^{3/2}/(3|F(t)|)$ denoting the
classical action of an electron returned back to the source region
along the unique closed orbit in a static electric field. Some
specific calculations are shown in Figs. 7 and 8 by varying the
single-cycle pulse duration and strength, respectively. It can be
found that the time-dependent calculations indeed approximate the
quasi-static results gradually when the field oscillation period
gets longer or the field strength becomes larger. The agreement
observed in Figs. 7 and 8 is best near the field peak position and
worst near the zero-field locations. However, the contribution in
the time-averaged observations mainly comes from the oscillations
near the field peak position because both the oscillation amplitude
and period near the zero-field locations are too small to give a
finite averaged signal, which confirms the validity of the static
field approximation used before. In Fig. 7(d), the returning-time
plot is given for the relevant closed orbits in Fig. 7(c), which
illustrates the physics behind the agreement observed in Figs. 7 and
8. Although the three types of closed orbits as in Fig. 2 are all
being there, only those closed orbits located near the line $t=t_i$
in Fig. 7(d) have an observable effect in the oscillation amplitude
of the photodetachment rate, because their durations $t-t_i$ are
much shorter than the others, and the associated returning-wave
amplitudes $\mathcal{A}$ in Eqs. (\ref{amplitude01}) and
(\ref{amplitude02}) are large enough.

\section{Conclusion}

Motivated by our recent studies on the temporal interferences in the
photodetachment of negative ions driven by a single-cycle THz
pulse\cite{BCYFR}, we examined the possible influences of a
single-cycle THz pulse on the time-dependent photodetachment rate.
We found that a weak THz pulse cannot change the total
photodetachment rate. However, if the applied THz pulse gets strong
enough, the photodetachment rate oscillates complicatedly. On the
other hand, we noticed that some classical trajectories of the
photoelectron can be driven back to the source region by a strong
single-cycle THz pulse. These observations remind us of a general
picture already recognized in the standard closed-orbit
theory\cite{COT01, COT02, COT03}, which addresses the correspondence
between the oscillatory photoionization (or photodetachemnt) rate
and the possible closed classical orbits embedded in the system.

To quantitatively understand the complex structures observed in the
time-dependent photodetachment rate, the standard closed-orbit
theory for the photodetachment in a static electric field has been
generalized to a time-dependent form which agrees well with exact
quantum simulations. The established formulas reveal a simple
dependence of the photodetachment-rate oscillations on the
properties of both the wave source and the closed classical orbits
existing in the system. Depending on the relative direction of the
returning orbit with respect to its initially-outgoing direction,
the photodetachment rates for different negative ions such as H$^-$
and F$^-$ might oscillate in phase or out of phase. In contrast to
the case of a static electric field\cite{COT_detachment, BCYDD}, the
oscillation amplitude of the photodetachment rate contributed by
each closed orbit has an additional term determined by the electron
returning momentum which is usually different from the
initially-outgoing momentum. As the applied electric field gets
stronger or its oscillation period becomes longer, the oscillatory
behavior of the photodetachment rate is more and more like that
obtained from the static-field approximation as in Refs.
\cite{microwave01, microwave02}.

The presented theory provides a clear and intuitive picture for the
photodetachment dynamics driven by a general time-dependent electric
field. Benefiting from the correlation between the electron launch
time and its later-returning time, a similar pump-probe technique as
in Ref. \cite{APT} may be a possible candidate in future experiments
for exploring the quantum effect of closed classical orbits from a
time-dependent viewpoint. An immediate application of the current
theory would be the photodetachment of negative ions in a static
electric field plus a strong oscillating electric field, where the
averaged photodetachment rate can be detected as in Ref.
\cite{RFfield}. For a weak oscillating field, the perturbation
formulas in Ref. \cite{Haggerty} can be used. More interesting
physics can be expected when the oscillating field amplitude is
comparable to or even larger than the static field strength, which
is also an interesting topic in future studies.

\begin{center}
{\bf ACKNOWLEDGMENTS}
\end{center}
\vskip8pt B. C. Y. thanks George Simion for helpful discussions.
This work was supported by the U. S. Department of Energy, Office of
Science, Basic Energy Sciences, under Award No. DE-SC0012193. This
research was supported in part through computational resources
provided by Information Technology at Purdue, West Lafayette,
Indiana.

\appendix

\section{THE DERIVATION OF EQS. (\ref{amplitude02}) and (\ref{extended_action_expression})}

The electron motion equation can be formally written as
\begin{equation}\label{A01}
   \text{$\rho=\rho(t_i,\theta_i, t)$ and $z=z(t_i,\theta_i, t)$}
\end{equation}
in the cylindric-coordinate frame. If we fix the electron final
destination ($\rho$, $z$) but let the other parameters change, the
following partial differential equation can be obtained,
\begin{equation}\label{A02}
    \bigg(\frac{\partial\rho}{\partial t}\bigg)_{t_i,\theta_i}\bigg(\frac{\partial t}{\partial
    t_i}\bigg)_{\rho, z}+\bigg(\frac{\partial\rho}{\partial \theta_i}\bigg)_{t_i,t}\bigg(\frac{\partial \theta_i}{\partial
    t_i}\bigg)_{\rho, z}=-\bigg(\frac{\partial\rho}{\partial t_i}\bigg)_{t,\theta_i}
\end{equation}
which can be explicitly written as
\begin{equation}\label{A03}
    \sin\theta_i\bigg(\frac{\partial t}{\partial
    t_i}\bigg)_{\rho, z}+(t-t_i)\cos\theta_i\bigg(\frac{\partial \theta_i}{\partial
    t_i}\bigg)_{\rho, z}=\sin\theta_i
\end{equation}
using Eq. (\ref{rho}). Similarly, we have
\begin{eqnarray}
\nonumber
   \big[k_0\cos\theta_i+A(t)-A(t_i)\big]\bigg(\frac{\partial t}{\partial
    t_i}\bigg)_{\rho, z}-k_0(t-t_i)\sin\theta_i&& \\
    \label{A04}
  \times\bigg(\frac{\partial \theta_i}{\partial
    t_i}\bigg)_{\rho, z}= k_0\cos\theta_i-F(t_i)(t-t_i)~~~&&
\end{eqnarray}
from Eq. (\ref{z}). After eliminating the partial derivative
$\big(\frac{\partial \theta_i}{\partial t_i}\big)_{\rho, z}$ in Eqs.
(\ref{A03}) and (\ref{A04}), we get
\begin{equation}\label{A05}
    \bigg(\frac{\partial t}{\partial
    t_i}\bigg)_{\rho,
    z}=\frac{k_0-F(t_i)(t-t_i)\cos\theta_i}{k_0+[A(t)-A(t_i)]\cos\theta_i}~.
\end{equation}
For the closed orbits returning back to the atom center, the partial
derivative $\big(\frac{\partial t}{\partial t_i}\big)_{\rho, z}$
becomes $dt/dt_i$, and Eq. (\ref{amplitude02}) is obtained by
combining Eqs. (\ref{A05}) and (\ref{amplitude01}) and also using
$p_z(t)=k_0\cos\theta_i+A(t)-A(t_i)$ with $\theta_i=0$ or $\pi$.

The classical action $\mathcal{S}$ along an arbitrary trajectory has
been obtained as\cite{BCYFR}
\begin{equation}\label{A06}
    \mathcal{S}=E_0(t-t_i)+z(t)\Delta p_z(t)-\frac{1}{2}\int_{t_i}^t[\Delta
    p_z(t')]^2dt'
\end{equation}
with the momentum transfer
\begin{equation}\label{A07}
    \Delta p_z(t)=A(t)-A(t_i)~.
\end{equation}
For the closed orbits, $z(t)=0$, and the action $\mathcal{S}$ can be
unfolded as
\begin{eqnarray}
\nonumber
  \mathcal{S} &=& E_0(t-t_i)-\frac{1}{2}A^2(t_i)(t-t_i) \\
\label{A08}
   && +A(t_i)\int_{t_i}^tA(t')dt'-\frac{1}{2}\int_{t_i}^t[A(t')]^2dt'
\end{eqnarray}
by substituting Eq. (\ref{A07}) into the last integration in Eq.
(\ref{A06}). The closed-orbit condition in Eq. (\ref{criteria})
allows the above equation to be further simplified as
\begin{equation}\label{A09}
    \mathcal{S}=\Big[E_0+\frac{1}{2}A^2(t_i)-A(t_i)k_0\cos\theta_i\Big](t-t_i)-\frac{1}{2}\int_{t_i}^tA^2(t')dt'
\end{equation}
which gives Eq. (\ref{extended_action_expression}) after the
definition in Eq. (\ref{extended_action}).

\section{THE DERIVATION OF EQ. (\ref{overlap_integration})}

To be clear, we write the oscillatory term $\Upsilon_\nu(t)$ in Eq.
(\ref{rate_adapt}) as the following form
\begin{equation}\label{B04}
    \Upsilon_{\pm z}(t)=-2f_L(t)f_L(t_i)\text{Im}\langle D\varphi_i|\widetilde{\psi}^{(\pm z)}_{\mathrm{ret}}\rangle
\end{equation}
where the notation $\pm z$ is used to indicate the returning
direction of the closed orbit. Specifically,
\begin{equation}\label{B05}
\widetilde{\psi}^{(+z)}_{\mathrm{ret}}=C(k_0)\mathcal{G}_{\mathrm{co}}Y_{lm}(\theta_i,
\phi_i)e^{ik_{\mathrm{ret}}z}
\end{equation}
representing the returning electron wave along the positive-$z$
direction, and
\begin{equation}\label{B06}
\widetilde{\psi}^{(-z)}_{\mathrm{ret}}=C(k_0)\mathcal{G}_{\mathrm{co}}Y_{lm}(\theta_i,
\phi_i)e^{-ik_{\mathrm{ret}}z}
\end{equation}
denoting the returning wave along the negative-$z$ direction. In
both Eq. (\ref{B05}) and Eq. (\ref{B06}), $\theta_i$ can be $0$ or
$\pi$, and the complex term $\mathcal{G}_{\mathrm{co}}$ is given by
Eq. (\ref{Gco_reduced}).

The overlap integration $\langle D\varphi_i|\widetilde{\psi}^{(\pm
z)}_{\mathrm{ret}}\rangle$ in Eq. (\ref{B04}) has almost the same
form as that studied in Ref. \cite{BCYDD} except that the electron
returning momentum $k_{\mathrm{ret}}$ is not conserved in our
current system. On the other hand, we note that the inhomogenous
Schr\"{o}dinger equation
\begin{equation}\label{B07}
    \Big(\frac{1}{2}\nabla^2+\frac{1}{2}k^2-V(r)\Big)\widetilde{\psi}^{(k)}_{\mathrm{out}}=D\varphi_i
\end{equation}
should be valid for any values of the momenta $k$, where the related
outgoing wave $\widetilde{\psi}^{(k)}_{\mathrm{out}}$ has the same
asymptotic form as in Eq. (\ref{outgoing}) but with a different
momentum value. Therefore, the same idea used in Appendix A of Ref.
\cite{BCYDD} can also be implemented, and the imaginary part of the
overlap integration in Eq. (\ref{B04}) can be converted to
\begin{equation}\label{B08}
    \text{Im}\langle D\varphi_i|\widetilde{\psi}_{\mathrm{ret}}\rangle=\frac{1}{2}
    \text{Im}\int\big(\widetilde{\psi}_{\mathrm{out}}\nabla_r\widetilde{\psi}^{*}_{\mathrm{ret}}-\widetilde{\psi}^{*}_{\mathrm{ret}}
    \nabla_r\widetilde{\psi}_{\mathrm{out}}\big)ds_{r}
\end{equation}
after replacing the source term $D\varphi_i$ by the outgoing wave
function $\widetilde{\psi}^{(k_{\mathrm{ret}})}_{\mathrm{out}}$ with
$k=k_{\mathrm{ret}}$ in Eq. (\ref{B07}). The integration in Eq.
(\ref{B08}) is on a spherical surface with a radius $r$ centered at
the negative ion.

Following Appendix B in Ref. \cite{BCYDD}, the two relevant
integrations in Eq. (\ref{B08}) can be worked out as,
\begin{eqnarray}
\nonumber
 && \int\widetilde{\psi}_{\mathrm{out}}\nabla_r\big[\widetilde{\psi}^{(\pm
    z)}_{\mathrm{ret}}\big]^*ds_{r} \\
\label{B14}
  &=& \frac{1}{2}(\pm
  1)^l\mathcal{C}_{lm}\sqrt{4\pi(2l+1)}\big[1+e^{i(2k_{\mathrm{ret}}r-l\pi)}\big]\delta_{m0}~~~~~~~~~
\end{eqnarray}
and
\begin{eqnarray}
\nonumber
 && \int\big[\widetilde{\psi}^{(\pm
    z)}_{\mathrm{ret}}\big]^*\nabla_r\widetilde{\psi}_{\mathrm{out}}ds_{r} \\
\label{B15}
  &=& -\frac{1}{2}(\pm
  1)^l\mathcal{C}_{lm}\sqrt{4\pi(2l+1)}\big[1-e^{i(2k_{\mathrm{ret}}r-l\pi)}\big]\delta_{m0},~~~~~~~~
\end{eqnarray}
where
\begin{equation}\label{B12}
    \mathcal{C}_{lm}=C(k_{\mathrm{ret}})C^*(k_0)\mathcal{G}^*_{\mathrm{co}}Y^*_{lm}(\theta_i,
\phi_i)~.
\end{equation}
The expression in Eq. (\ref{overlap_integration}) is obtained
by substituting Eqs. (\ref{B14}) and (\ref{B15}) into Eq.
(\ref{B08}) with Eqs. (\ref{B12}) and (\ref{Ylm}) together. Note
that the $r$-dependent terms in Eqs. (\ref{B14}) and (\ref{B15})
cancel each other.



\begin{references}

\bibitem{Gutzwiller}
M. C. Gutzwiller, \textit{Chaos in Classical and Quantum Mechanics}
(Springer-Verlag, New York, 1990).

\bibitem{review}
D. Kleppner and J. B. Delos, Found. Phys. {\bf 31}, 593 (2001) and
references therein.

\bibitem{COT01}
M. L. Du and J. B. Delos, Phys. Rev. Lett. {\bf 58}, 1731 (1987).

\bibitem{COT02}
M. L. Du and J. B. Delos, Phys. Rev. A {\bf 38}, 1896 (1988).

\bibitem{COT03}
M. L. Du and J. B. Delos, Phys. Rev. A {\bf 38}, 1913 (1988).

\bibitem{JGao01}
J. Gao, J. B. Delos, and M. Baruch, Phys. Rev. A {\bf 46}, 1449
(1992).

\bibitem{JGao02}
J. Gao and J. B. Delos, Phys. Rev. A {\bf 46}, 1455 (1992).

\bibitem{Peters01}
A. D. Peters and J. B. Delos, Phys. Rev. A {\bf 47}, 3020 (1993).

\bibitem{Peters02}
A. D. Peters, C. Jaff\'{e}, and J. B. Delos, Phys. Rev. A {\bf 56},
331 (1997).

\bibitem{Main01}
J. Main, G. Wiebusch, K. Welge, J. Shaw, and J. B. Delos, Phys. Rev.
A {\bf 49}, 847 (1994).

\bibitem{Main02}
J. Main and G. Wunner, Phys. Rev. A {\bf 55}, 1743 (1997).

\bibitem{Matzkin}
A. Matzkin, P. A. Dando, and T. S. Monteiro, Phys. Rev. A {\bf 66},
013410 (2002).

\bibitem{Wright}
J. D. Wright, J. M. DiSciacca, J. M. Lambert, and T. J. Morgan,
Phys. Rev. A {\bf 81}, 063409 (2010).

\bibitem{COT_detachment}
M. L. Du, Phys. Rev. A {\bf 70} 055402 (2004).

\bibitem{BCYDD}
B. C. Yang, J. B. Delos, and M. L. Du, Phys. Rev. A {\bf 89}, 013417
(2014).

\bibitem{RFfield}
N. Spellmeyer, D. Kleppner, M. R. Haggerty, V. Kondratovich, J. B.
Delos, and J. Gao, Phys. Rev. Lett. {\bf 79}, 1650 (1997).

\bibitem{Haggerty}
M. R. Haggerty and J. B. Delos, Phys. Rev. A {\bf 61}, 053406
(2000).

\bibitem{Rydberg01}
S. Li and R. R. Jones, Phys. Rev. Lett. {\bf 112}, 143006 (2014).

\bibitem{Rydberg02}
B. C. Yang and F. Robicheaux, Phys. Rev. A {\bf 90}, 063413 (2014).

\bibitem{Rydberg03}
B. C. Yang and F. Robicheaux, Phys. Rev. A {\bf 91}, 043407 (2015).

\bibitem{Nelson}
S. Fleischer, Y. Zhou, R. W. Field, and K. A. Nelson, Phys. Rev.
Lett. {\bf 107}, 163603 (2011).

\bibitem{Bob}
K. N. Egodapitiya, S. Li, and R. R. Jones, Phys. Rev. Lett. {\bf
112}, 103002 (2014).

\bibitem{BCYFR}
B. C. Yang and F. Robicheaux, Phys. Rev. A {\bf 92}, 063410 (2015).

\bibitem{Fabrikant}
I. I. Fabrikant, Sov. Phys.-JETP {\bf{52}}, 1045 (1980).

\bibitem{Demkov}
Y. N. Demkov, V. D. Kondratovich and V. N. Ostrovskii, JETP Lett.
{\bf{34}}, 403 (1981).

\bibitem{Du4983}
M. L. Du, Phys. Rev. A {\bf 40}, 4983 (1989).

\bibitem{Blondel}
C. Blondel, C. Delsart and F. Dulieu, Phys. Rev. Lett. {\bf 77},
3755 (1996).

\bibitem{microwave01}
M. C. Baruch, T. F. Gallagher, and D. J. Larson, Phys. Rev. Lett.
{\bf 65}, 1336 (1990).

\bibitem{microwave02}
M. C. Baruch, W. G. Sturrus, N. D. Gibson, and D. J. Larson, Phys.
Rev. A {\bf 45}, 2825 (1992).

\bibitem{microwave03}
A. Bugacov, B. Piraux, M. Pont, and R. Shakeshaft, Phys. Rev. A {\bf
45}, 3041 (1992).

\bibitem{microwave04}
S. Bivona, R. Burlon, and C. Leone, Phys. Rev. A {\bf 45}, 3268
(1992).

\bibitem{Chu}
C. Laughlin and Shih-I Chu, Phys. Rev. A {\bf 48}, 4654 (1993).

\bibitem{Lin}
X. X. Zhou, Z. J. Chen, T. Morishita, A. T. Le, and C. D. Lin, Phys.
Rev. A {\bf 77}, 053410 (2008).

\bibitem{Rost01}
A. K\"{a}stner, U. Saalmann, and J. M. Rost, Phys. Rev. Lett. {\bf
108}, 033201 (2012).

\bibitem{Rost02}
A. K\"{a}stner, U. Saalmann, and J. M. Rost, J. Phys. B: At. Mol.
Opt. Phys. {\bf 45}, 074011 (2012).

\bibitem{Bracher_exact}
C. Bracher, T. Kramer and M. Kleber, Phys. Rev. A {\bf 67}, 043601
(2003).

\bibitem{Bracher_thesis}
C. Bracher, Ph. D. thesis, Technische Universit\"{a}t M\"{u}nchen,
1999.

\bibitem{Wigner}
E. P. Wigner, Phys. Rev. {\bf 73}, 1002 (1948).

\bibitem{Du5609}
M. L. Du and J. B. Delos, Phys. Rev. A {\bf 38}, 5609 (1988).

\bibitem{BCYbarrier}
B. .C Yang and M. L. Du, J. Phys. B: At. Mol. Opt. Phys. {\bf 45},
175003 (2012).

\bibitem{APT}
T. Remetter, P. Johnsson, J. Mauritsson, K. Varj\'{u}, Y. Ni, F.
L\'{e}pine, E. Gustefsson, M. Kling, J. Khan, R. L\'{o}pez-Martens,
K. J. Schafer, M. J. J. Vrakking and A. L'huillier, Nature Physics
{\bf 2}, 323 (2006).


\end{references}
\end{document}